\documentclass[a4paper]{jpconf}
\usepackage{graphicx}
\usepackage{xspace}

\newcommand{\pp}{\mbox{$p+p$}\xspace}
\newcommand{\auau}{\mbox{Au$+$Au}\xspace}
\newcommand{\pT}{\mbox{$p_{\rm T}$}\xspace}
\newcommand{\dpTpT}{\mbox{$\delta\pT/\pT$}\xspace}
\newcommand{\raa}{\mbox{$R_{\rm AA}$}\xspace}
\newcommand{\taa}{\mbox{$T_{\rm AA}$}\xspace}

\newcommand{\Npart}{\mbox{$N_{\rm part}$}\xspace}
\newcommand{\Nqp}{\mbox{$N_{\rm qp}$}\xspace}

\newcommand{\dndeta}{\mbox{$dN_{\rm ch}/d\eta$}\xspace}
\newcommand{\sloss}{\mbox{$S_{\rm loss}$}\xspace}

\newcommand{\snn}{\mbox{$\sqrt{s_{_{\rm NN}}}$}\xspace}

\newcommand{\piz}{\mbox{$\pi^0$}\xspace}

\begin{document}
\title{Detailed study of parton energy loss via measurement of 
fractional momentum loss of high \pT hadrons in heavy ion collisions}

\author{Takao Sakaguchi, for the PHENIX Collaboration}

\address{Brookhaven National Laboratory, Physics Department, Upton, NY 11973-5000, USA}

\ead{takao@bnl.gov}

\begin{abstract}
PHENIX measurement of the fractional momentum loss (\dpTpT) of
high \pT identified hadrons are presented. The \dpTpT of high \pT \piz
which are computed from 39\,GeV Au+Au over to 2.76\,TeV Pb+Pb are found
to vary by a factor of six. We plotted the \dpTpT against several
global variables, \Npart, \Nqp and \dndeta, and found global features.
It was found that 200\,GeV Au+Au points are merging into the central
2.76\,TeV Pb+Pb points when plotting \dpTpT against \dndeta.
\end{abstract}

\section{Introduction}
\begin{figure}[h]
\begin{minipage}{70mm}
\centering
\includegraphics[width=6.5cm]{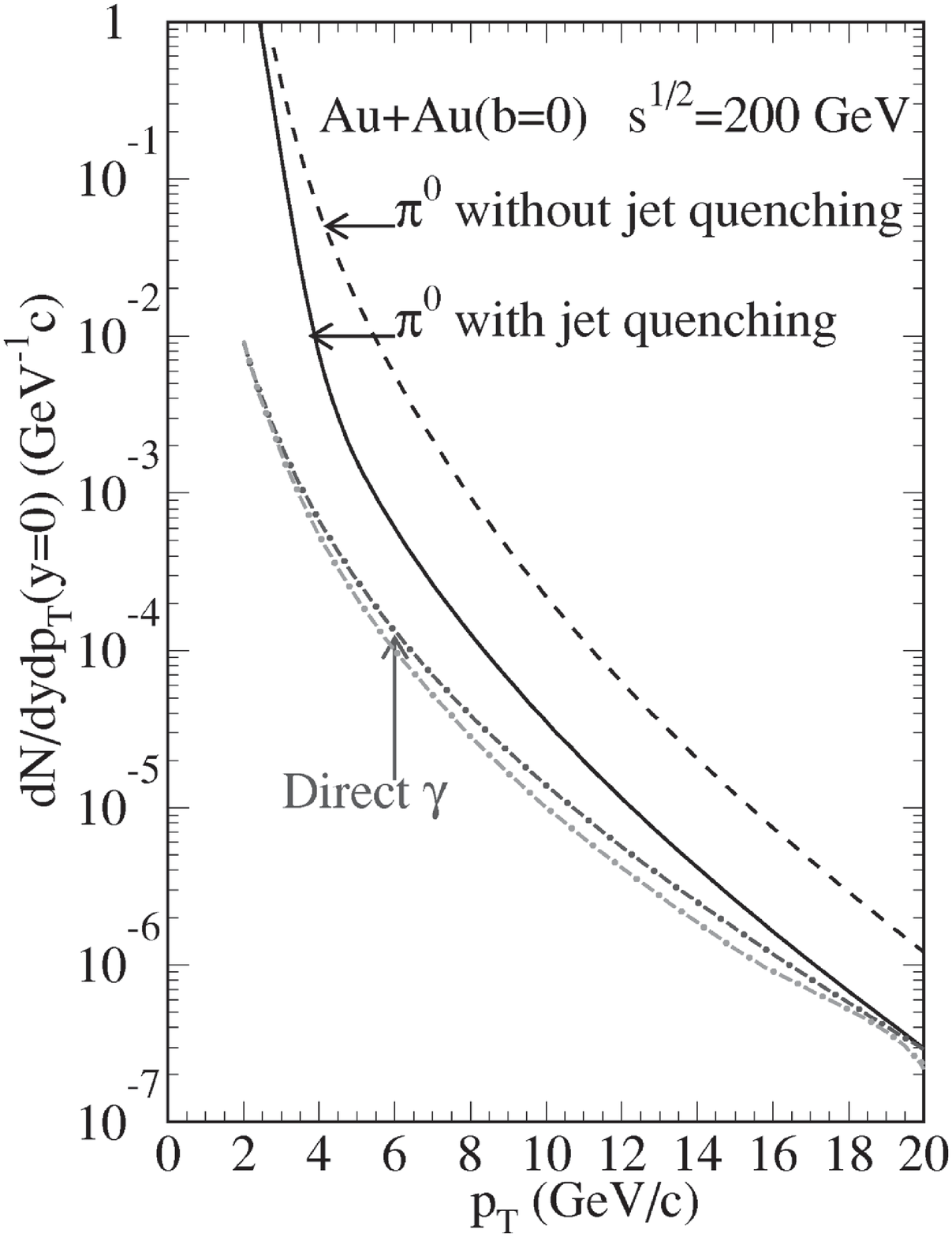}
\end{minipage}
\begin{minipage}{90mm}
\centering
\includegraphics[width=8.5cm]{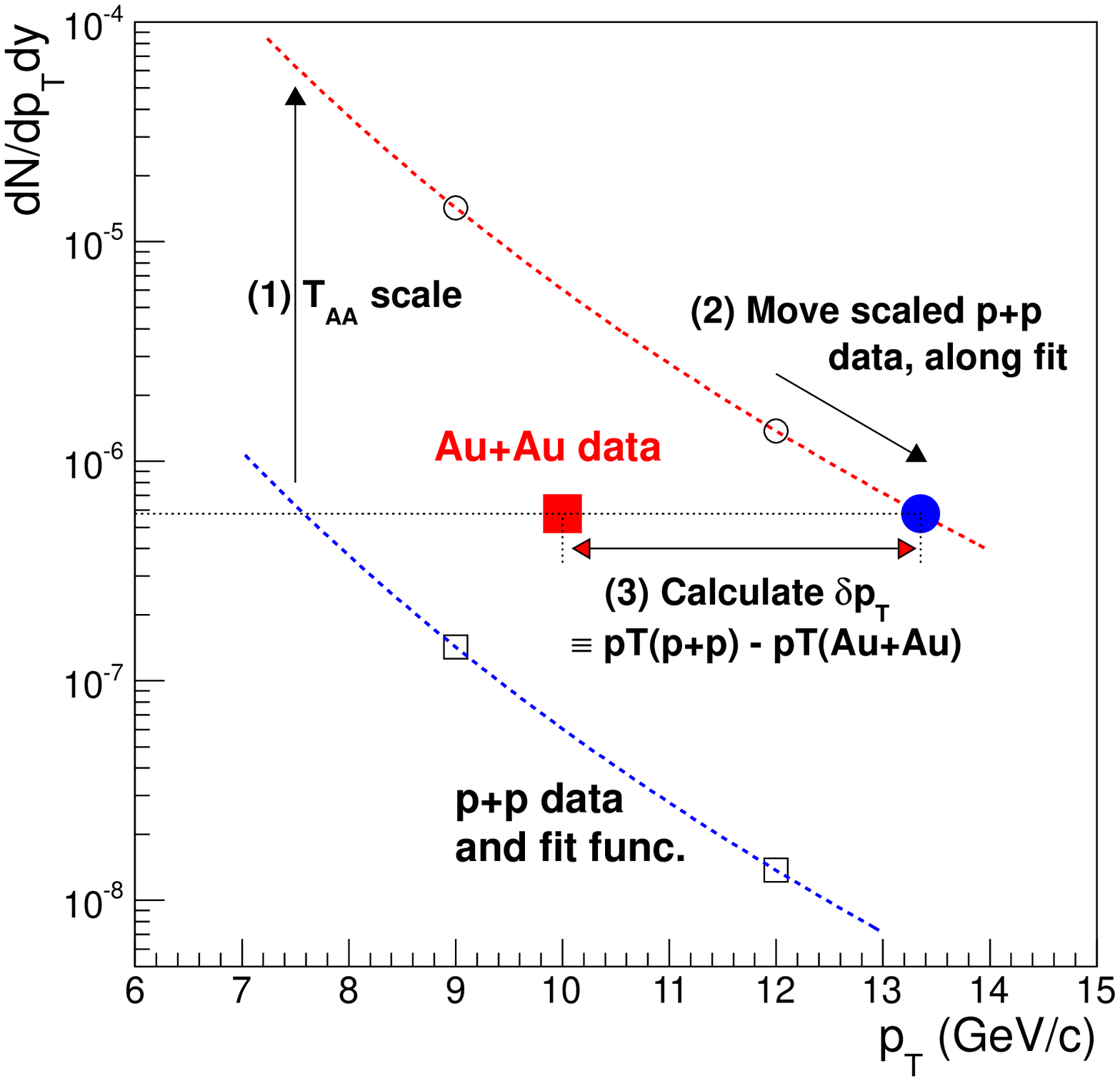}
\end{minipage}
  \caption{(a, left) A calculation demonstrating that the suppression of the
\piz yield is resulted in from the shift of the momentum
spectra in A+A collisions. (b, right) Method of calculating average \dpTpT.
 We scaled the $p+p$ yield by $T_{\rm AA}$ corresponding to centrality
 selection of Au+Au data, shifted the $p+p$ points closest to Au+Au in yield,
 and calculated the momentum difference of $p+p$ and Au+Au points.
  }
\label{fig1}
\end{figure}

The interaction of hard scattered partons with the medium created by heavy ion
collisions (i.e., quark-gluon plasma, QGP) has been of interest since the
beginning of the RHIC running~\cite{Wang:1998bha}. A large suppression of the
yields of high transverse momentum (\pT) hadrons which are the fragments of
such partons was observed, suggesting that the matter is sufficiently dense
to cause parton-energy loss prior to hadronization~\cite{Adler:2003qi}.
The PHENIX experiment~\cite{Adcox:2003zm} has been exploring the highest
\pT region with single \piz mesons which are leading hadrons of jets.
We show a calculation on the energy loss of partons published almost
20 years ago in Fig.~\ref{fig1}(a).
Although the measurement of the momentum shift is the ultimate goal,
the paper suggested looking at
the ratio of the high \pT single hadrons in Au+Au and p+p collisions
as an alternate way. Since then, most of the experiments including
PHENIX have looked at the nuclear modification factors,
\raa ($\equiv(dN_{\rm AA}/dyd\pT)/(\langle T_{\rm AA}\rangle d\sigma_{pp}/dyd\pT)$),
and quantified the energy loss effect via its suppression.
We here present the momentum shift of high \pT hadrons instead of \raa.

\section{Fractional momentum loss \dpTpT}
\begin{figure}[h]
\begin{minipage}{74mm}
\centering
\vspace{5mm}
\includegraphics[width=7.3cm]{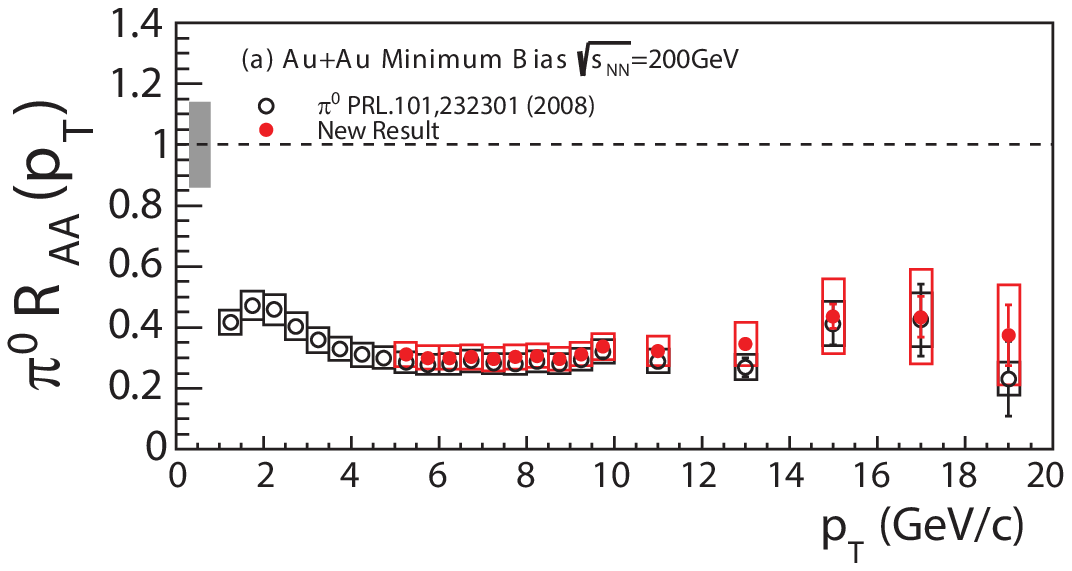}
\end{minipage}
\begin{minipage}{85mm}
\centering
\includegraphics[width=8.3cm]{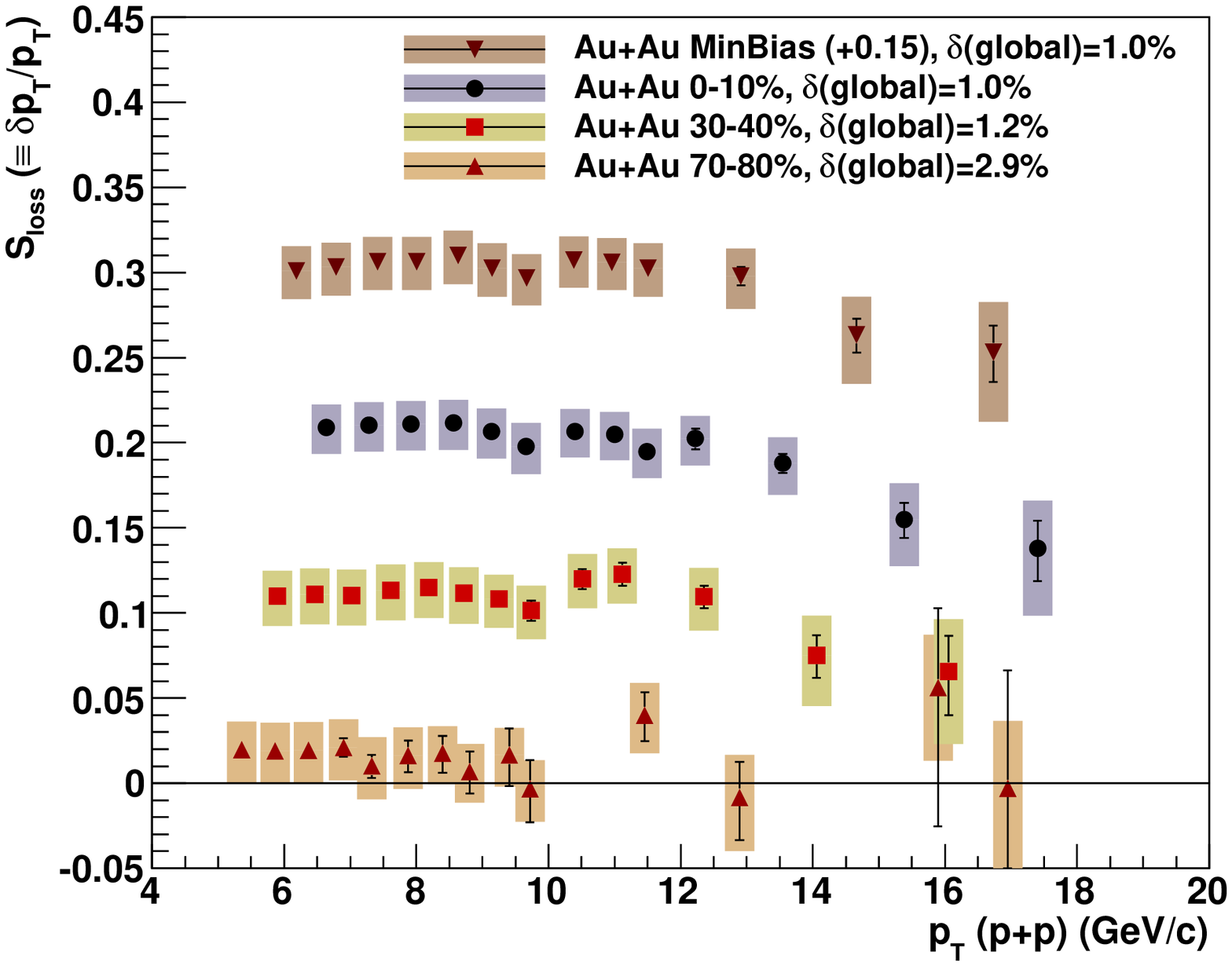}
\end{minipage}
\caption{(a, left) $R_{AA}$ of $\pi^0$'s for 200\,GeV Au+Au collisions obtained from RHIC Year-7 run. (b, right) \dpTpT for the same dataset.} 
\label{fig2}
\end{figure}

\begin{figure}[h]
\begin{minipage}{68mm}
\centering
\vspace{5mm}
\includegraphics[width=6.8cm]{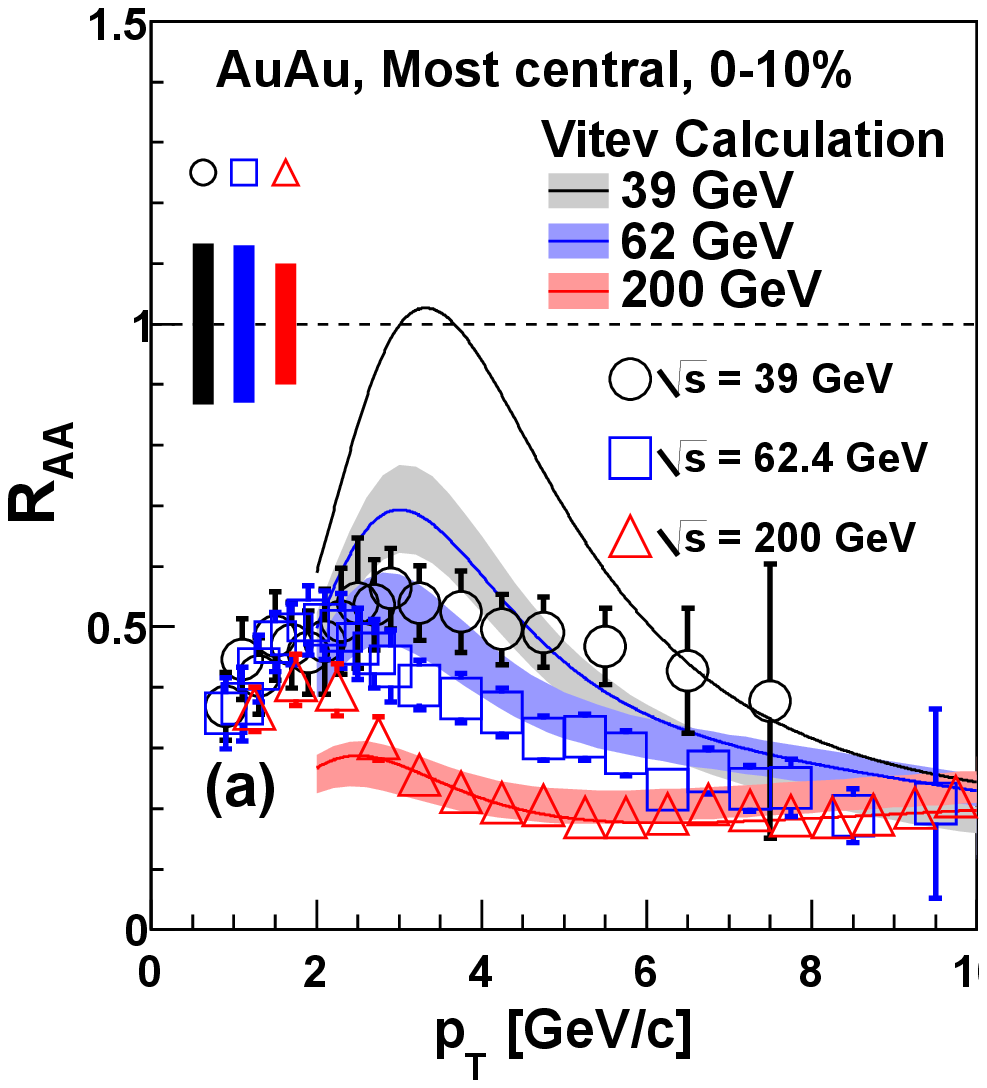}
\end{minipage}
\begin{minipage}{85mm}
\centering
\includegraphics[width=6.8cm]{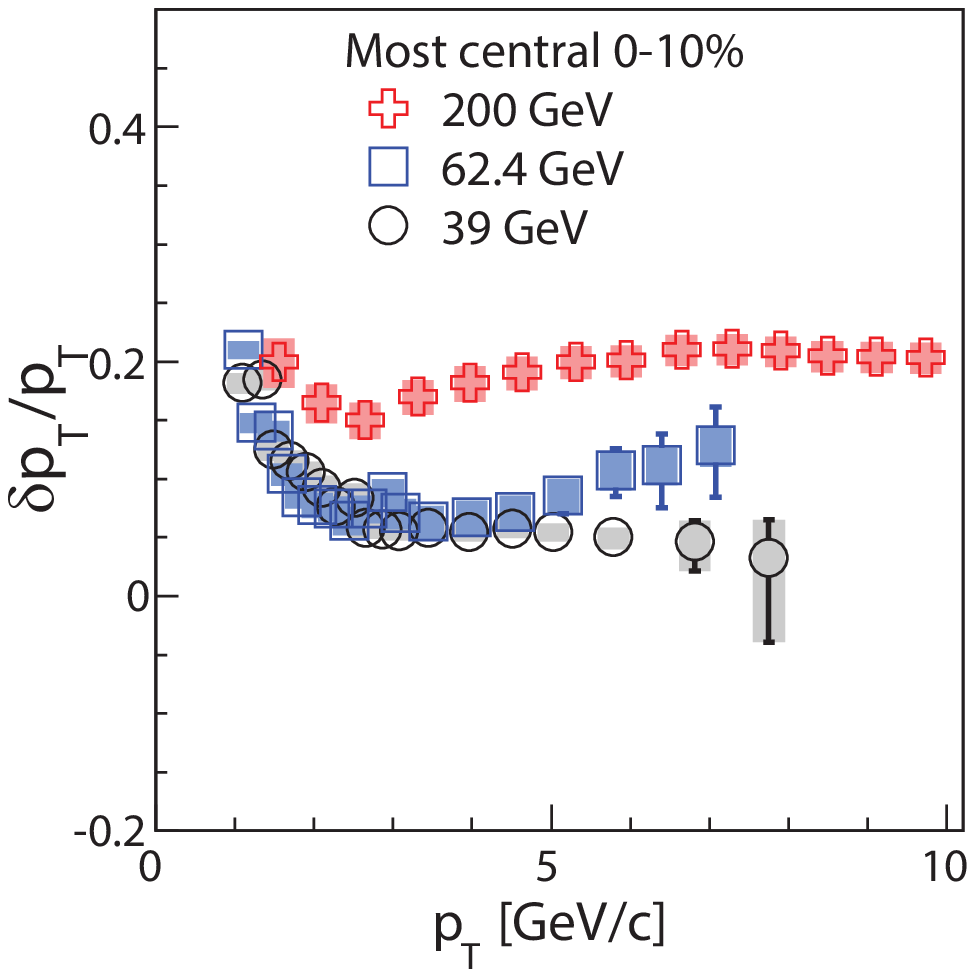}
\end{minipage}
\caption{(a, left) $R_{AA}$ of $\pi^0$'s for 39, 62 in 0-10\,\% Au+Au collisions obtained from RHIC Year-10 run and in 0-10\,\% 200\,GeV Au+Au collisions from RHIC Year-10 run. (b, right) \dpTpT for the same dataset.} 
\label{fig3}
\end{figure}

With a larger statistics of both \pp and \auau data recently collected,
it became possible to measure the momentum shift directly.
Fig.~\ref{fig1}(b) depicts the method to compute such shift.
We have statistically extracted the fractional momentum loss
 ($\dpTpT$, $\delta \pT \equiv \pT - \pT'$, where \pT is the transverse
momentum of the $p+p$ data, and \pT' is that of the Au+Au data) of the
partons using the hadron \pT spectra measured in $p+p$ and Au+Au
collisions~\cite{Adare:2012wg}.
Since the number of data points is finite, a fit to the scaled \pp is
needed to evaluate \dpTpT at a given \auau invariant yield.
The uncertainty of the \dpTpT is calculated by inversely converting
the quadratic sum of the uncertainties on the yields of \auau and \pp
points, using the \pp fit function. Statistical and systematic
uncertainties are individually calculated in the same way. The
uncertainties on \taa and p+p luminosity are not plotted but mentioned
in plots.
Fig.~\ref{fig2}(a) show the $R_{AA}$ for the $\pi^0$'s in Au+Au collisions
at \snn=200\,GeV from the RHIC Year-7 run. Using the \dpTpT calculation
method, we obtained the \dpTpT for the same dataset as shown in
Fig.~\ref{fig2}(b).
Similarly, the $R_{AA}$ for the $\pi^0$'s in 0-10\,\% Au+Au
collisions at \snn=39, 62 and 200\,GeV from the RHIC Year-7 and Year-10
runs shown in Fig.~\ref{fig3}(a) are replotted in the form of \dpTpT
as shown in Fig.~\ref{fig3}(b).
The $R_{AA}$'s look similar even the cms energies are changed by a factor
of 5 as seen in Fig.~\ref{fig3}(a). However, when we look at \dpTpT for the
corresponding dataset, we found that the \dpTpT changes by a factor of
three from 39 to 200\,GeV as shown in Fig.~\ref{fig3}(b)~\cite{Adare:2012uk}.

\begin{figure}[h]
\begin{minipage}{68mm}
\centering
\vspace{5mm}
\includegraphics[width=6.8cm]{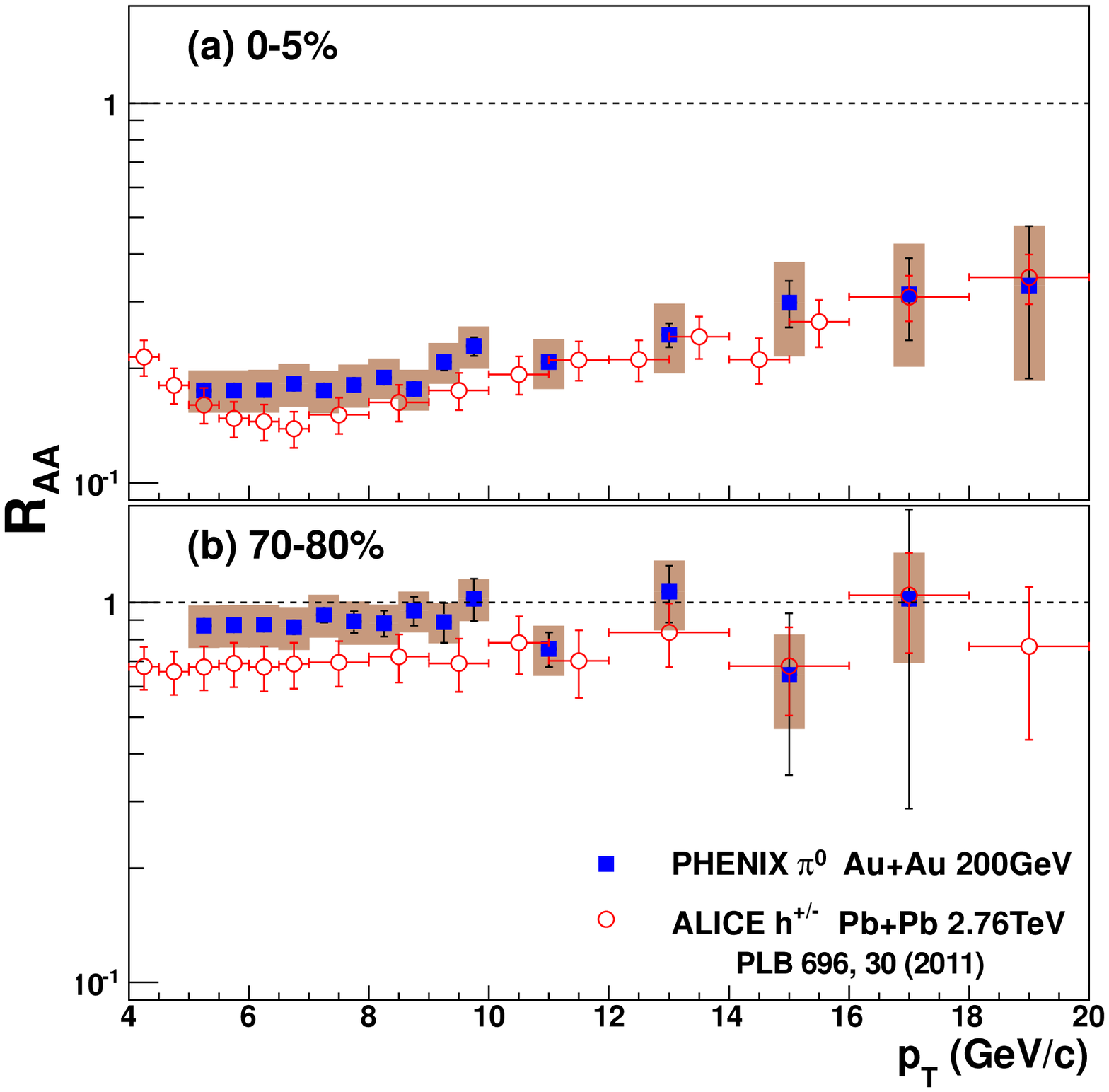}
\end{minipage}
\begin{minipage}{85mm}
\centering
\includegraphics[width=8.3cm]{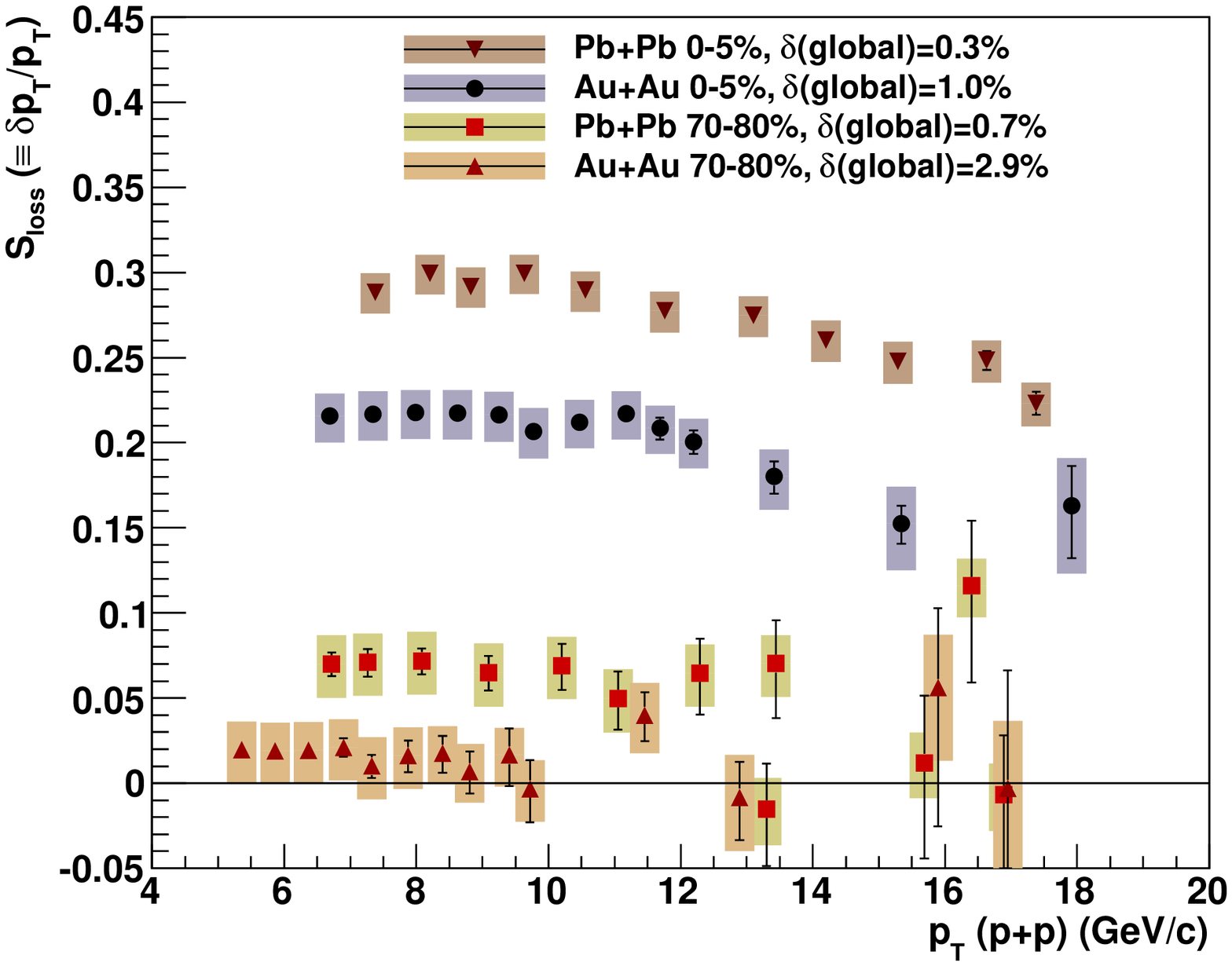}
\end{minipage}
\caption{(a, left) $R_{AA}$ of $\pi^0$'s for 200\,GeV Au+Au collisions obtained from RHIC Year-7 run and charged hadrons for 2.76\,TeV Pb+Pb collisions obtained by the ALICE experiment at LHC~\cite{Aamodt:2010jd}. (b, right) \dpTpT for the same dataset.} 
\label{fig4}
\end{figure}

The $R_{AA}$'s also look similar between RHIC and LHC (Fig.~\ref{fig4}(a)).
Similarly, the \dpTpT is found to change by a factor of $\sim1.5$ from
200 to 2.76\,TeV (Fig.~\ref{fig4}(b)).
To summarize, even the $R_{AA}$'s are similar, the \dpTpT's show a factor
of six variation from 39\,GeV to 2.76\,TeV. This fact has not been found
by looking at $R_{AA}$.

\section{Scaling property of \dpTpT}
\begin{figure}[h]
\centering
\includegraphics[width=0.8\linewidth]{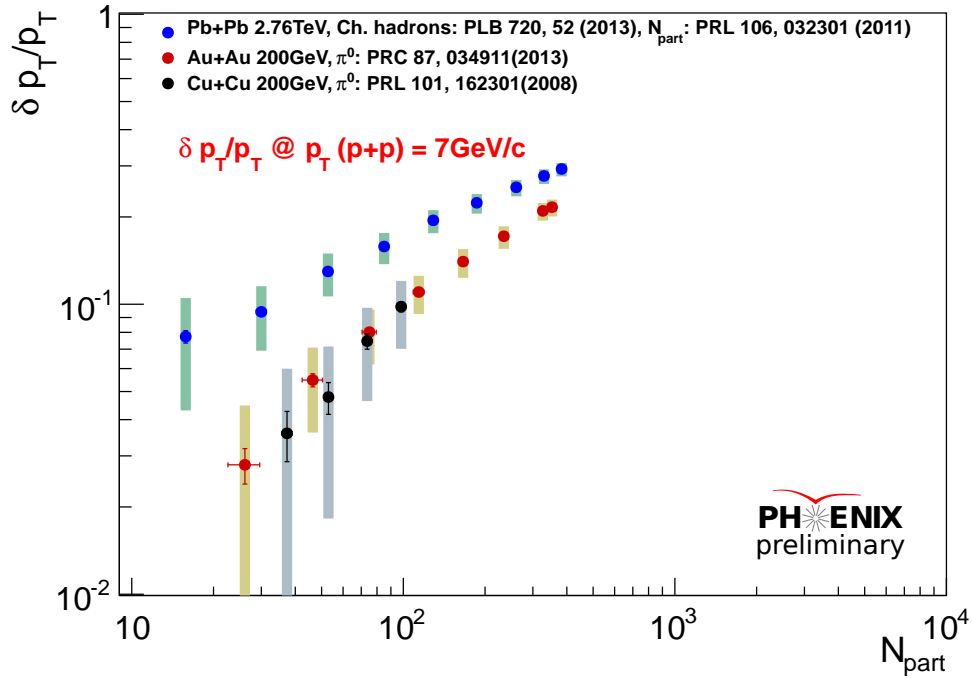}
\caption{\dpTpT as a function of \Npart for $\pi^0$'s in 200\,GeV Au+Au and Cu+Cu collisions measured by PHENIX and charged hadrons in 2.76\,TeV Pb+Pb collisions measured by ALICE.}
\label{fig5}
\end{figure}

\begin{figure}[h]
\centering
\includegraphics[width=0.8\linewidth]{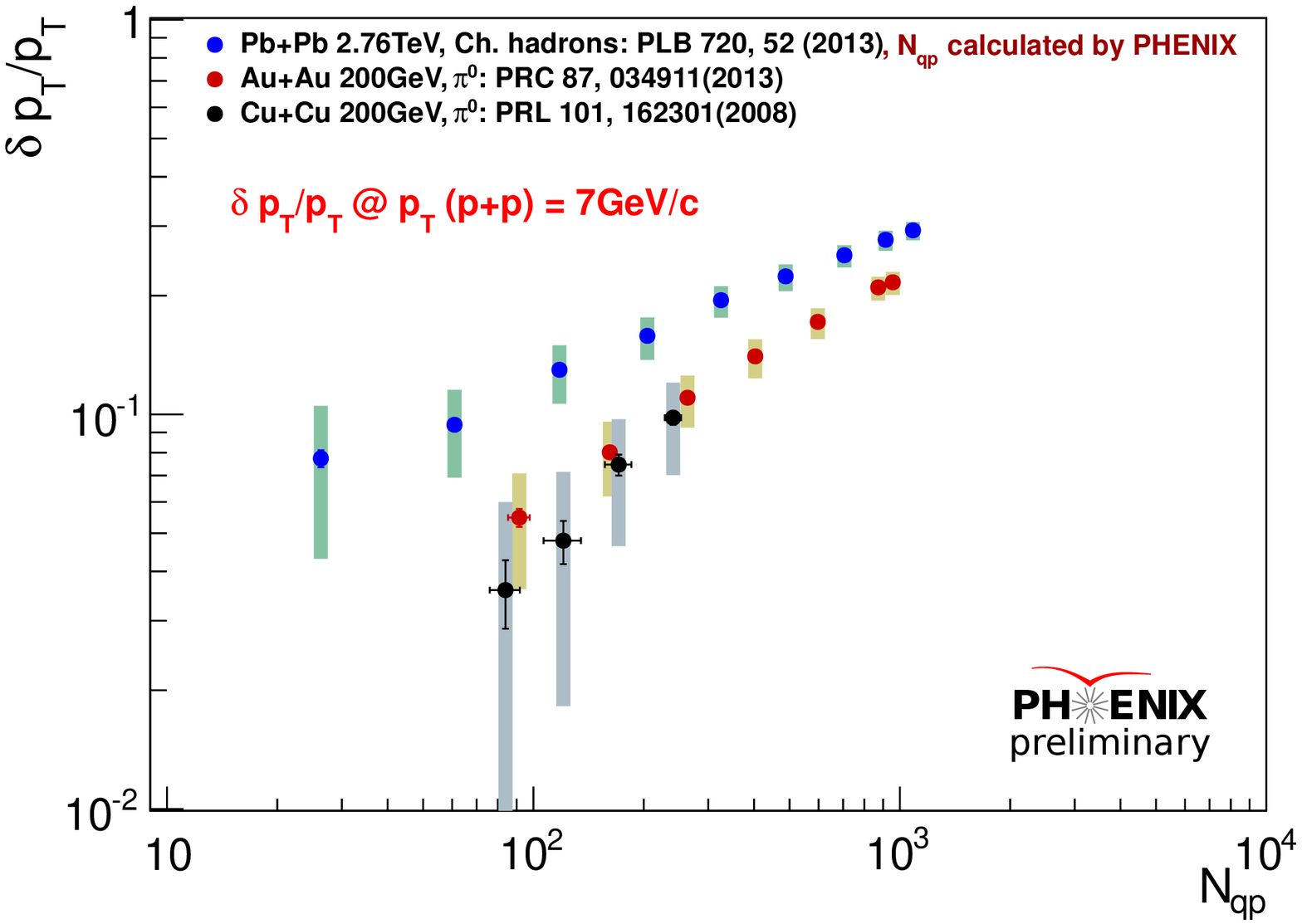}
\caption{\dpTpT as a function of \Nqp for $\pi^0$'s in 200\,GeV Au+Au and Cu+Cu collisions measured by PHENIX and charged hadrons in 2.76\,TeV Pb+Pb collisions measured by ALICE.}
\label{fig6}
\end{figure}

\begin{figure}[h]
\centering
\includegraphics[width=0.8\linewidth]{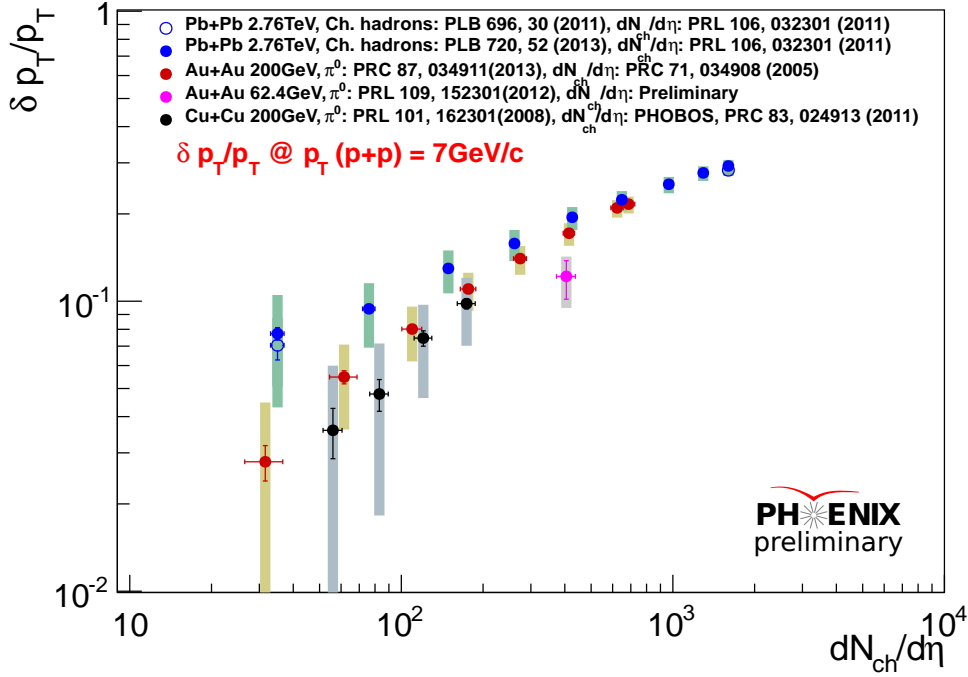}
\caption{\dpTpT as a function of \dndeta for $\pi^0$'s in Au+Au collisions at 200\,GeV and 62.4\,GeV, and in Cu+Cu collisions at 200\,GeV measured by PHENIX and charged hadrons in 2.76\,TeV Pb+Pb collisions measured by ALICE.}
\label{fig7}
\end{figure}

In order to study the systematics of \dpTpT, we plot the \dpTpT against
several global variables such as \Npart, \Nqp (number of quark
participants) and \dndeta. We first plotted the \dpTpT against \Npart
as shown in Fig.~\ref{fig5}.
All the plots shown in this section are at $\pT(p+p) =7$\,GeV/c in
order to reach the hard scattering regime. In the \Npart scaling
we see that the Cu+Cu and Au+Au are nicely lined up, implying that within
the same cms energy, the \dpTpT scales with \Npart. This is consistent
with the fact that $R_{AA}$ is similar at same \Npart between Cu+Cu and
Au+Au collisions~\cite{Adare:2008ad}. The Pb+Pb points are consistently
off the trend of 200\,GeV points, but the slopes of both systems look
similar. Fig.~\ref{fig6} shows \dpTpT against \Nqp.
The detail description of how the number of quark participants
are obtained can be found in the literature~\cite{Adler:2013aqf}.
We employed a Monte-Carlo-Glauber (MC-Glauber) model to calculate
the numbers. We first determine the quark-quark inelastic cross section
($\sigma^{\rm inel}_{qq}$) for each collision energy such that the
inelastic nucleon-nucleon cross section ($\sigma^{\rm inel}_{NN}$)
is reproduced. Then the model is modified to handle the quark-quark
rather than nucleon-nucleon collisions.
The nuclei are placed according to a Woods-Saxon distribution and
then three quarks are distributed around the center of each nucleon
following the distribution of:
\[ \rho^{proton}(r) = \rho^{proton}_{0} \times e^{-ar}\]
where $a = \sqrt{12}/r_{m} = 4.27$\,fm$^{-1}$ and $r_{m}=0.81$\,fm
is the rms charge radius of the proton.
A pair of quarks, one from each nucleus, interact with each other
if their distance $d$ in the plane transverse to the beam axis 
satisfies the condition of $d<\sqrt{\sigma^{\rm inel}_{qq}/\pi}$.
The number of quark participants as a function of the number of 
nucleon participants is nonlinear, especially for low values of \Npart. 
In Fig.~\ref{fig6}, the similar feature as the previous plot is
seen. Since the \Npart is a factor of 2-3 higher than \Nqp, all
the points are systematically moved to the right.
Finally, we plotted the \dpTpT against the charged multiplicity,
\dndeta, as shown in Fig.~\ref{fig7}.
In this plot, we added one 62\,GeV Au+Au point which is 0-10\,\% centrality.
We expect that \dndeta well represents the energy density of the system.
It is interesting to note that the most central Au+Au 200\,GeV points tend
to merge into the most central points of Pb+Pb collisions, while they
deviate each other as going to lower \dndeta. This systematic trend has
not been found by looking at \raa's which look similar across the systems.
In order to cross-check this new result, we have performed a power-law
fit to \dpTpT vs \dndeta points from 200\,GeV Au+Au collisions, and
compared the power with the result obtained from a different
method~\cite{Adare:2008qa}. We fitted the points of this work with
$\dpTpT = \beta (\dndeta)^{\alpha/1.19}$ assuming
\dndeta$\propto$(\Npart)$^{1.19}$~\cite{Adler:2004zn}, and obtained
$\alpha$ as 0.64$\pm$0.07.
Assuming the spectra shape follows a power-law with the power $n$, one can
write the relation between \dpTpT and \raa as:
\[ \sloss \equiv \dpTpT = \beta \Npart^{\alpha}, \\
\raa = (1-\sloss)^{n-2} = (1-\beta \Npart^{\alpha})^{n-2} \]
Following this relation, we obtained the power $\alpha$ as 0.57$\pm$0.13 from
the fit to the integrated \raa as a function of \Npart in the
literature~\cite{Adare:2008qa}. We therefore confirmed that the powers
obtained by the two methods are consistent.

\section{Summary}
We presented PHENIX measurement of the fractional momentum loss (\dpTpT)
of high \pT identified hadrons. By looking at
the \dpTpT instead of \raa, we found many interesting features.
The \dpTpT of high \pT \piz which are computed from 39\,GeV Au+Au
over to 2.76\,TeV Pb+Pb are found to vary by a factor of six.
We plotted the \dpTpT against several global variables, \Npart, \Nqp and
\dndeta. It was found that 200\,GeV Au+Au points are merging into the
central 2.76\,TeV Pb+Pb points when plotting \dpTpT against \dndeta.
We performed a power-law fit to the \dpTpT vs \dndeta, and
obtained a power that is consistent with the one obtained from the fit to
the integrated \raa.
We are going to add points from other systems to systematically
investigate the \dpTpT.












\end{document}